\documentclass[twocolumn]{aastex61}

\usepackage{multirow}

\newcommand{\gpy}{Gpc$^{-3}$\,yr$^{-1}$}

\submitjournal{ApJ}

\shorttitle{The Origin of r-Process Elements}
\shortauthors{C\^ot\'e et al.}
\begin{document}

\title{The Origin of r-Process Elements in the Milky Way}

\correspondingauthor{Benoit C\^ot\'e}
\email{bcote@uvic.ca, benoit.cote@csfk.mta.hu}

\author[0000-0002-9986-8816]{Benoit C\^ot\'e}
\affil{Konkoly Observatory, Research Centre for Astronomy and Earth Sciences, Hungarian Academy of Sciences, Konkoly Thege Miklos ut 15-17, H-1121 Budapest, Hungary}
\affiliation{Joint Institute for Nuclear Astrophysics - Center for the Evolution of the Elements, USA}
\affiliation{NuGrid Collaboration, \href{http://nugridstars.org}{http://nugridstars.org}}

\author[0000-0003-2624-0056]{Chris L. Fryer}
\affiliation{Center for Theoretical Astrophysics, LANL, Los Alamos, NM, 87545, USA}
\affiliation{Joint Institute for Nuclear Astrophysics - Center for the Evolution of the Elements, USA}
\affiliation{NuGrid Collaboration, \href{http://nugridstars.org}{http://nugridstars.org}}

\author[0000-0002-1658-7681]{Krzysztof Belczynski}
\affiliation{Nicolaus Copernicus Astronomical Center, Polish Academy of Sciences, ul. Bartycka 18, 00-716 Warsaw, Poland}

\author[0000-0003-4156-5342]{Oleg Korobkin}
\affiliation{Center for Theoretical Astrophysics, LANL, Los Alamos, NM, 87545, USA}
\affiliation{Joint Institute for Nuclear Astrophysics - Center for the Evolution of the Elements, USA}

\author{Martyna Chru\'sli\'nska}
\affiliation{Institute of Mathematics, Astrophysics and Particle Physics, Radboud University Nijmegen, PO box 9010, 6500 GL Nijmegen, the Netherlands}

\author{Nicole Vassh}
\affiliation{University of Notre Dame, Notre Dame, Indiana 46556, USA}

\author[0000-0002-9950-9688]{Matthew R. Mumpower}
\affiliation{Center for Theoretical Astrophysics, LANL, Los Alamos, NM, 87545, USA}
\affiliation{Theoretical Division, Los Alamos National Lab, Los Alamos, NM, 87545, USA}
\affiliation{Joint Institute for Nuclear Astrophysics - Center for the Evolution of the Elements, USA}

\author[0000-0002-5936-3485]{Jonas Lippuner}
\affiliation{Center for Theoretical Astrophysics, LANL, Los Alamos, NM, 87545, USA}
\affiliation{Joint Institute for Nuclear Astrophysics - Center for the Evolution of the Elements, USA}

\author{Trevor M. Sprouse}
\affiliation{University of Notre Dame, Notre Dame, Indiana 46556, USA}

\author{Rebecca Surman}
\affiliation{University of Notre Dame, Notre Dame, Indiana 46556, USA}
\affiliation{Joint Institute for Nuclear Astrophysics - Center for the Evolution of the Elements, USA}

\author[0000-0003-3265-4079]{Ryan Wollaeger}
\affiliation{Center for Theoretical Astrophysics, LANL, Los Alamos, NM, 87545, USA}



\begin{abstract}

Some of the heavy elements, such as gold and europium (Eu), are almost exclusively formed by the rapid neutron capture process (r-process).  However, it is still unclear which astrophysical site between core-collapse supernovae and neutron star - neutron star (NS-NS) mergers produced most of the r-process elements in the universe.  Galactic chemical evolution (GCE) models can test these scenarios by quantifying the frequency and yields required to reproduce the amount of europium (Eu) observed in galaxies. Although NS-NS mergers have become popular candidates, their required frequency (or rate) needs to be consistent with that obtained from gravitational wave measurements. Here we address the first NS-NS merger detected by LIGO/Virgo (GW170817) and its associated Gamma-ray burst and analyze their implication on the origin of r-process elements. The range of NS-NS merger rate densities of $320-4740$\,\gpy\ provided by LIGO/Virgo is remarkably consistent with the range required by GCE to explain the Eu abundances in the Milky Way with NS-NS mergers, assuming the solar r-process abundance pattern for the ejecta. Under the same assumption, this event has produced about $1-5$ Earth masses of Eu, and $3-13$ Earth masses of gold. When using theoretical calculations to derive Eu yields, constraining the role of NS-NS mergers becomes more challenging because of nuclear astrophysics uncertainties. This is the first study that directly combines nuclear physics uncertainties with GCE calculations. If GW170817 is a representative event, NS-NS mergers can produce Eu in sufficient amounts and are likely to be the main r-process site.

\end{abstract}




%%%%%%%%%%%%%%%%%%%
%%%%%%%%%%%%%%%%%%%
\section{Introduction} \label{sec:intro}
%%%%%%%%%%%%%%%%%%%
%%%%%%%%%%%%%%%%%%%

Core-collapse supernovae (CC~SNe) and neutron star - neutron star (NS-NS) mergers are the two leading candidates for producing most of the rapid neutron capture process (r-process) elements in the universe (e.g., \citealt{2004A&A...416..997A,arnould07,2014MNRAS.438.2177M,2015A&A...577A.139C,2015MNRAS.452.1970W}).  NS-NS mergers, originally proposed by \cite{lattimer74}, recently gained popularity because the high neutron fraction allows robust production of the 2nd and 3rd r-process peaks (e.g., \citealt{freiburghaus99,korobkin12,bauswein13a,2016MNRAS.460.3255R,2017ARNPS..6701916T}, but see \citealt{2015ApJ...810..109N}). If NS-NS mergers are indeed more likely to produce the full r-process, the challenge is now to determine whether the rate of NS-NS mergers is high enough to explain the r-process enrichment observed in the Milky Way and other galaxies.

In \citet[C17a]{cote17a}, we derived the rate of NS-NS mergers required in galactic chemical evolution (GCE) studies in order to match the amount of europium (Eu) observed in the Milky Way, assuming NS-NS mergers are the dominant r-process site.  Eu in the solar system is almost entirely made by the r-process (\citealt{2000ApJ...544..302B}) and is therefore used as a tracer. The GCE studies compiled in C17a (which include our own study) cover a wide range of numerical approaches from one-zone models to cosmological hydrodynamic simulations (\citealt{2014MNRAS.438.2177M,2015A&A...577A.139C,2015ApJ...814...41H,2015ApJ...804L..35I,2015ApJ...807..115S,2015MNRAS.452.1970W,2016ApJ...830...76K}, see also \citealt{2015MNRAS.447..140V} and \citealt{2017arXiv170703401N}). Within the uncertainties, we found that the required merger rates can be consistent with the upper limits provided by Advanced LIGO during their first observing run  (\citealt{2016ApJ...832L..21A}), although they are systematically higher than the rates predicted by the population synthesis models of \cite{2016Natur.534..512B}.

One of the goals of our previous work was to create a bridge between the GCE community and future LIGO/Virgo detections. In this paper, we apply our methodology to address the first NS-NS merger ever detected via gravitational waves (GW170817, \citealt{abbott17}), which provides new estimates for the NS-NS merger rate density in the nearby universe.  This merger manifested itself across the entire electromagnetic spectrum from radio through gamma-rays, providing additional constraints on the location, distance, and the ejecta mass and composition~\citep{cowperthwaite17,evans17,tanvir17,troja17}. We define the abundance pattern of GW170817 based on light curve fits of the ultraviolet (UV), optical, and infrared (IR) emissions, calculate the impact of nuclear physics uncertainties on the r-process yields (which was not done in C17a), and include those uncertainties in a GCE context. We also update the population synthesis predictions seen in C17a using the latest models of \cite{Chruslinska17}.

This paper is organized as follows. We present in Section~\ref{sect_yields} the NS-NS merger yields derived from the multi-wavelength observations of GW170817 and discuss the impact of nuclear physics uncertainties.  In Section~\ref{sect_methods}, we tie our GCE and population synthesis predictions with LIGO/Virgo's rate and yield measurements. We discuss the implication of this new detection in Section~\ref{sect_r_MW}, and conclude in Section~\ref{sect_concl}.

%%%%%%%%%%%%%%%%%%%%%
%%%%%%%%%%%%%%%%%%%%%
%%%%%%%%%%%%%%%%%%%%%
\section{Merger Yields} \label{sect_yields}
%%%%%%%%%%%%%%%%%%%%%
%%%%%%%%%%%%%%%%%%%%%
%%%%%%%%%%%%%%%%%%%%%

The ejecta from NS-NS mergers can be classified into two main categories which are distinguished by the time of ejection: dynamical ejecta, generated at the time of contact, and everything else which emerges after the single object is formed, broadly referred to as ``wind" ejecta either from a disk or a hypermassive neutron star \citep{metzger12,metzger16c}.  
Estimates of the dynamical ejecta mass in various theoretical models vary from $10^{-4}\,M_\odot$ to 0.1\,$M_\odot$ 
(see e.g. \citealt{hotokezaka13}, \citealt{sekiguchi16a}, \citealt{lehner16a}, \citealt{bovard17}, \citealt{cote17a}, and \citealt{dietrich17} for reviews).  The dynamical ejecta is expelled so fast that it preserves very low electron fraction ($Y_e<0.2$, \citealt{rosswog13}), leading to the robust production of the so-called ``main" r-process from the 2nd to 3rd r-process peaks (e.g., Figure~3 of \citealt{wollaeger17}).  However, general relativistic simulations which include neutrino irradiation predict a broader distribution of $Y_e$ with a tail which extends over 0.3 \citep[e.g.][]{bovard17}. A distribution such as this covers the entire r-process range from the 1st peak all the way to the 3rd \citep{wanajo14,tanaka17a}.  

Estimates for the masses in the ``wind'' category of the ejecta vary from $10^{-4}\,M_\odot$ up to a few $10^{-1}\,M_\odot$ \citep{fernandez13,perego14a,fernandez15,Just2015,cote17a,siegel17}. The electron fraction distribution, and hence the composition, of the wind ejecta is also uncertain. However, the general consensus is that the wind ejecta should have a higher electron fraction ($Y_e = 0.2-0.5$) than the dynamical ejecta, thus producing isotopes in the range between the 1st and 2nd r-process peaks, or even near the iron peak for particularly high $Y_e$ values \citep{lippuner15,lippuner17b,martin15}.

Until recently, attempts to detect kilonova were limited to observations of nearby gamma-ray bursts, placing only weak constraints on the ejecta mass and composition. The infrared excess in GRB~130603B~\citep{tanvir13} suggested $\sim$\,0.05\,$M_\odot$ of neutron rich ejecta. But the accompanying bump in X-ray emission points instead to an afterglow flare origin for the infrared excess, arguing for a lower ejecta mass for the neutron rich material. Other studies found $\sim0.1\ M_\odot$ in the case of GRB~050714~\citep{yang15} and similarly high mass for GRB~060614~\citep{jin15}, which due to the uncertainties could still be treated as strict upper limits on the ejecta mass. Upper limits from GRB~160821B are more strict, suggesting that at least some bursts have less than $0.01-0.03$\,$M_\odot$ of neutron rich ejecta \citep{kasliwal17a}.

%%%%%%%%%%%%%%%%%%%%%%%%%%%%%%%%%%%%%%%%%%%%
%%%%%%%%%%%%%%%%%%%%%%%%%%%%%%%%%%%%%%%%%%%%
\subsection{Ejecta From GW170817}
\label{sec:ejecta}
%%%%%%%%%%%%%%%%%%%%%%%%%%%%%%%%%%%%%%%%%%%%
%%%%%%%%%%%%%%%%%%%%%%%%%%%%%%%%%%%%%%%%%%%%

Fits to the UV, optical, and IR data from GW170817 provide 
definitive constraints on the masses and general composition of ejecta components.  
Table~\ref{tab:ejmasses} summarizes the relevant findings in recent literature.
All studies agree that at least two spatially separated ejecta components were
present: low-opacity radioactive material to power the early optical emission
\citep{nicholl17,evans17}, and high-opacity lanthanide-polluted outflow to account 
for late-time near-IR emission \citep{arcavi17b,troja17,tanvir17}.
Note that  only the latter contains Eu, 
while the former component is lanthanide-free and expected to have a relatively high $Y_e$.

Studies with bolometric light curve reconstruction
\citep{smartt17,kasliwal17b,cowperthwaite17,rosswog17b} provide simple and robust 
lower limits on the total mass of the ejecta ($m_{\rm total}>0.03-0.05\ M_\odot$) and upper limits on the neutron richness, measured in terms of electron fraction: 
$Y_e\lesssim0.3$ \citep{rosswog17b}. The latter constraint follows 
from the shape of the nuclear heating profile as reflected in bolometric light curve.

\begin{deluxetable}{lcc}
\tablewidth{0pc}
\tablecaption{Estimates of ejected masses for high-opacity lanthanide-rich material ($m_{\rm dyn}$) 
and medium-opacity ``winds'' ($m_{\rm w}$), harvested from the recent literature for GW170817. \label{tab:ejmasses}}
\tablehead{\colhead{Reference} & \colhead{$m_{\rm dyn}\,[M_\odot]$} 
                               & \colhead{$m_{\rm w}\,[M_\odot]$} }      
\startdata
\cite{abbott17c}       & $0.001-0.01$   &   --             \\  
\cite{arcavi17b}       &  --            & $0.02-0.025$     \\  
\cite{cowperthwaite17} & $0.04$         &   $0.01$         \\  
\cite{chornock17}      & $0.035$        &   $0.02$         \\
\cite{evans17}         & $0.002-0.03$   &  $0.03-0.1$      \\  
\cite{kasen17}         & $0.04$         &   $0.025$        \\
\cite{kasliwal17b}     & $>0.02$        &  $>0.03$         \\
\cite{nicholl17}       & $0.03$         &   --             \\
\cite{perego17}        & $0.005-0.01$   &  $10^{-5}-0.024$ \\
\cite{rosswog17b}      & $0.01$         &   $0.03$         \\
\cite{smartt17}        & $0.03-0.05$    &   $0.018$        \\
\cite{tanaka17b}       & $0.01$         &   $0.03$         \\
\cite{tanvir17}        & $0.002-0.01$   &   $0.015$        \\  
\cite{troja17}         & $0.001-0.01$   &  $0.015-0.03$    \\  
\enddata
\end{deluxetable}

\begin{deluxetable*}{lcccccc}
\tablewidth{0pc}
\tabletypesize{\footnotesize}
\tablecaption{Yields of the r-process constituents. 
Second column: mass fractions $X^{\rm obs}$ in the solar r-process residuals~\citep{arnould07}.  
Third column: mass fractions in the hypothetical low-$Y_e$ dynamical ejecta when adopting the r-process residuals abundance pattern for $A>110$.
Fourth and fifth columns: mass fractions with theoretical nucleosynthetic yields with low-$Y_e$ and high-$Y_e$ ejecta, respectively.
The two remaining columns show the ejected masses inferred for GW170817: 
$M_{\rm ejected}^{\rm obs}$ -- using the observed r-process residuals, 
$M_{\rm ejected}^{\rm w/ \, nuc.}$ -- using theoretical yields. 
$A$ and $Z$ represent the mass and atomic numbers, respectively. \label{tab:yields}}
\tablehead{ \colhead{Abundance} & \colhead{$X^{\rm obs}$ [$10^{-8}$]} 
  &\colhead{$X^{\rm obs}_{{\rm low}\,Y_e}$} 
  &\colhead{$X^{\rm w/nuc.}_{{\rm low}\,Y_e}$} 
  &\colhead{$X^{\rm w/nuc.}_{{\rm high}\,Y_e}$} 
  &\colhead{$M_{\rm ejected}^{\rm obs}$ [$M_\odot$]} 
  &\colhead{$M_{\rm ejected}^{\rm w/ \, nuc.}$ [$M_\odot$]}}
\startdata
Total r-process ($A>79$) &$35.0^{+0.4}_{-0.3}$   &$0.99^{\rm (a)}$          &${0.98-1.0}^{\rm (a)}$ &$0.56 - 0.70$ & $0.01-0.04$ & $0.0075-0.03$\\
Main r-process ($A>130$) &$7.08^{+0.09}_{-0.03}$ &$0.284^{+0.003}_{-0.003}$ &$0.70-0.99$            &$0.002-0.06$  & $(6-30)\times10^{-4}$ & $0.0014-0.012$ \vspace{0.03 in}\\
\hline \vspace{-0.075 in} \\
1st peak ($30<Z<38$)     &$7.92^{+0.09}_{-0.04}$ & 0.0                      &$0.0-2\times10^{-4}$   &$0.7 - 0.9$   & $0.004-0.012$ & $0.007-0.03$\\
2nd peak ($48<Z<59$)     &$3.50^{+0.07}_{-0.05}$ &$0.141^{+0.002}_{-0.002}$ &$0.2-0.8$              &$0.007 - 0.1$ & $0.004-0.013$ & $0.0005-0.011$\\    
3rd peak ($74<Z<83$)     &$1.62^{+0.20}_{-0.03}$ &$0.065^{+0.006}_{-0.001}$ &$0.13-0.5$             &$0.0$ & $(1.3-7)\times10^{-4}$ & $(0.3-5)\times10^{-3}$\\
Trans-lead ($Z>82$)   &$0.03^{+0.06}_{-0.02}$ &$1.2^{+1.6}_{-0.6}\times 10^{-3}$ &$0.006-0.154$ & $0.0$ &$(2-30)\times10^{-6}$ & $(0.12-15)\times10^{-4}$\\
Iron Peak ($21<Z<30$)    &0.0 &0.0 &0.0 &$0.0002-0.01$ &0.0 &$(5-14)\times10^{-6}$ \\
Europium ($Z=63$)   &$0.036^{+0.005}_{-0.001}$ &$1.45^{+0.14}_{-0.003}\times 10^{-3}$ & $0.0002-0.02$ &0.0 & $(3-15)\times10^{-6}$ & $(0.4-22)\times10^{-6}$\\
$^{56}$Ni &0.0 &0.0 &0.0 &0.0 &0.0 &0.0 \\
\enddata
\tablenotetext{a}{The remaining $\approx1\%$ of mass is theorized to be in ${}^4$He formed by $\alpha$-decays.}
\end{deluxetable*}        

For better agreement with observations, some studies argued that 
an additional third ejecta component is required, such as a wide-angle 
mildly relativistic cocoon 
\citep{kasliwal17b}, secular disk winds \citep{perego17} or a ``purple
kilonova'' outflow \citep{cowperthwaite17}. All these additional outflows have
relatively high electron fraction and do not produce Eu.
In a different approach, an axisymmetric two-component model with toroidal 
dynamical ejecta and spherical wind was applied \citep{wollaeger17}. Geometric 
effects in this model allow photon reprocessing and thus higher luminosities for 
the same masses, correspondingly producing lower mass estimates (see 
Table~\ref{tab:ejmasses}). 
In particular, based on this model, \cite{tanvir17} and \cite{troja17} 
required a moderate amount of high-$Y_e$ wind ejecta ($\sim0.015$\,$M_\odot$), 
with a relatively low amount of low-$Y_e$ dynamical ejecta 
($\sim0.002-0.01$\,$M_\odot$). The statistical MCMC analysis of large set of 
gray-opacity models in \cite{cowperthwaite17} and \cite{perego17}, on the other hand, 
argued for a total ejecta mass of 0.04\,$M_\odot$ with 1\% of this mass in lanthanides.
The mass of low-$Y_e$ component can also be estimated independently from fits
of synthetic spectra at late times 
\citep{kasliwal17b,chornock17,tanaka17b}.

%%%%%%%%%%%%%%%%%%%%%%%%%%%%%%%%%%%%%
%%%%%%%%%%%%%%%%%%%%%%%%%%%%%%%%%%%%%
\subsection{Adopted r-Process Yields}
\label{sec:adopted_ab_pat}
%%%%%%%%%%%%%%%%%%%%%%%%%%%%%%%%%%%%%
%%%%%%%%%%%%%%%%%%%%%%%%%%%%%%%%%%%%%

In this paper, we test the hypothesis that GW170817 is a typical, representative event which produces a regular r-process signature, consistent with the robust abundance pattern observed in metal-poor halo stars \citep{sneden08} and the ultra-faint dwarf galaxy Reticulum~II \citep{ji16,2016AJ....151...82R}. This case is studied in Section~\ref{sect_r_MW}. The other possibility of GW170817 being an unusual, non-representative case is discussed in Section~\ref{disc_unrep_event}, where theoretical r-process yields and their uncertainties (see Section~\ref{sec:uncertainties}) are applied. From these abundance patterns, we extract Eu yields in order to calibrate the NS-NS merger rates required by GCE studies, as they typically use Eu to trace the r-process enrichment (see Section~\ref{disc_using_Eu}).

Based on the estimates presented in Table~\ref{tab:ejmasses}, we adopt a conservatively broad range of $0.002-{0.01\,M_\odot}$ for dynamical ejecta, and $0.01-{0.03\,M_\odot}$ for the high-$Y_e$ component, which adds up to $0.01-0.04\,M_\odot$ for the total ejecta mass range. 
Table~\ref{tab:yields} summarizes the resulting yields of r-process constituents. 
The second column displays the observed composition of r-process residuals in the Solar system \citep[from][]{arnould07}. 
The third column contains a hypothetical normalized composition expected if the low-$Y_e$ component robustly produces the observed pattern \citep{korobkin12,bauswein13a,MT2015} starting from second peak on ($A>110$). 
Fourth and fifth columns contain theoretical compositions for low-$Y_e$ and high-$Y_e$ components, computed from first principles with nucleosynthesis network (see the following Section \ref{sec:uncertainties} for details). 
In the last two columns, we
convolve the ejecta mass ranges inferred from GW170817 with the observed or theoretical mass fractions of different constituents. The second to last column ({\small $M_{\rm ejected}^{\rm obs}$}) lists masses computed assuming the observed solar r-process residuals, while the last column ({\small $M_{\rm ejected}^{\rm w/ \, nuc.}$}) lists the ranges computed theoretically. 

Note that the high-$Y_e$ ``wind'' ejecta observed in GW170817 does not contain lanthanides and thus is not expected to produce 3rd peak r-process elements. In principle, the ratio of 3rd to 2nd r-process peaks from GW170817 may not match the solar abundance pattern. The uncertainties in wind versus dynamical ejecta masses for current light curve fits to GW170817 shown in Table~\ref{tab:ejmasses} allow for a wide range of ratios of the r-process peaks (see Section~\ref{disc_using_Eu} for discussion). However, this does not come in tension with the robust abundance pattern observed in the r-process-rich Galactic halo stars and in Reticulum II~\citep{ji16}, because the robustness only holds for $Z\ge56$ \citep[see e.g. Figure~11 in][]{sneden08}. It is currently unclear whether the robustness can be extended to the 2nd r-process peak, since atomic lines in the region $50\le Z\le55$ are in the UV range, which is notoriously hard to measure \citep[e.g.][]{roederer12,2012ApJ...750...76R}. In fact, some fission models of the robust r-process nucleosynthesis in the neutron-rich ejecta strongly underproduce the 2nd peak \citep{goriely13}.

%%%%%%%%%%%%%%%%%%%%%%%%%%%%%%%%%%%%
\subsection{Uncertainties from Nuclear Physics} \label{sec:uncertainties}
%%%%%%%%%%%%%%%%%%%%%%%%%%%%%%%%%%%%

The consistency between the r-process pattern observed in metal-poor halo stars \citep{sneden08}, the solar r-process residuals \citep{arnould07}, and the ultra-faint dwarf galaxy Reticulum~II \citep{ji16,2016AJ....151...82R} places tight constraints on the site of the main r-process. To agree with observations, a feasible candidate needs to be able to reproduce the observed pattern robustly, with little sensitivity to the variations in system parameters~\citep{korobkin12}. However, the current nuclear theory of heavy element nucleosynthesis produces uncertainties in the predicted pattern exceeding observational constraints by at least one order of magnitude \citep{mumpower2016r}. In this section, we explore the approach of extracting the r-process yields of GW170817 using nucleosynthesis calculations from first principles.
%For this reason, we explore two approaches to extract the r-process yields of GW170817: one in which the observed solar abundance distribution is assumed, and one in which we simulate the nucleosynthesis from first principles.

The nuclear physics uncertainties mainly stem from the fact that the r-process path meanders through the uncharted territory of heavy, extremely neutron-rich nuclei close to the neutron drip line, for which no experimental data is available. Variations in the unknown nuclear masses \citep{mumpower2016r}, fission fragment distribution~\citep[e.g.][]{goriely13}, neutron capture rates~\citep{mumpower12}, $\beta$-decay rates~\citep{eichler15,mumpower14}, and the specifics of the fission mechanism itself can all significantly impact nucleosynthetic yields \citep{mumpower2016r}. For this reason, we calculate abundance yields with a variety of nuclear mass models and fission fragment distributions. 

Mass model uncertainties are of utmost concern for nucleosynthesis calculations as they influence all major features of the r-process abundance pattern. In particular, the strength of the shell closures predicted by each model directly determines the shape of the primary r-process peaks while more subtle trends for highly deformed nuclei influence the rare-earth peak. We select ten popular mass models (DZ, FRDM1995, FRDM2012, HFB17, HFB21, HFB24, WS3, KTUY, SLY4, and UNEDF0) which are based on a wide range of physical underpinnings.

\begin{figure}
\center
\includegraphics[width=3.35in]{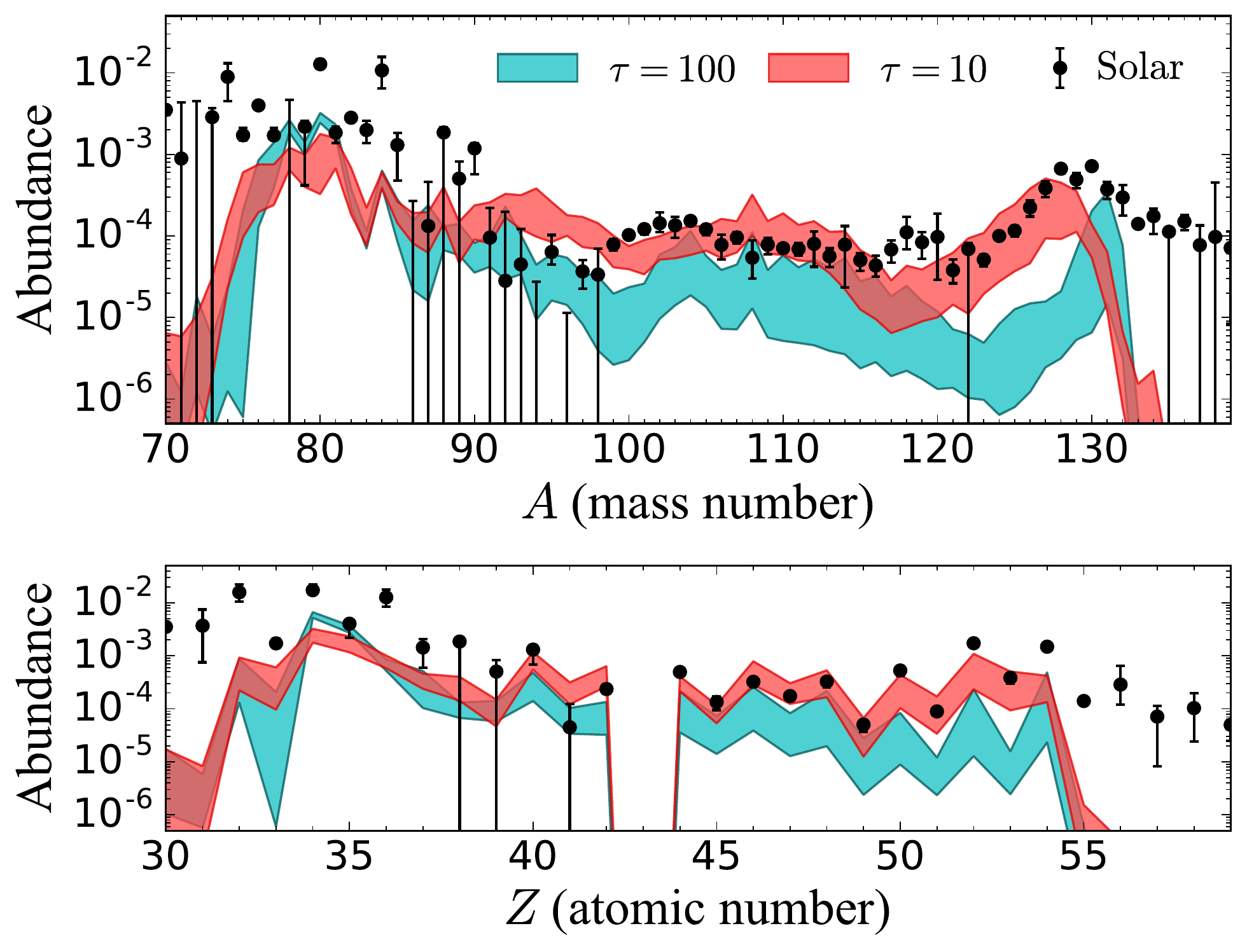}
\caption{The range in the calculated high-$Y_e$ ($Y_e=0.27$) r-process abundances as a function of mass number $A$ (upper panel) and atomic number $Z$ (lower panel) generated from a set of ten different mass models (DZ, FRDM1995, FRDM2012, HFB17, HFB21, HFB24, WS3, KTUY, SLY4, and UNEDF0). Turquoise bands represent low entropy wind conditions ($s=10$) with slow outflow while red bands represent similar conditions with a faster outflow timescale. The dots are the observed solar r-process residuals \citep[taken from][]{arnould07}. \label{fig:Yep27}}
\end{figure}

Some models, such as FRDM and WS3 are based on the macroscopic-microscopic approach which separates the macroscopic shape degrees of freedom from the microscopic description of the nucleus. Other models, such as SLY4 or UNEDF0 provide a fully microscopic description of the nucleus based on Skyrme interactions and density functional theory. The predictions of nuclear masses between these models tends to diverge as one approaches the neutron dripline resulting in a range of abundance patterns in calculations of nucleosynthesis.

\begin{figure*}
\includegraphics[width=7.0in]{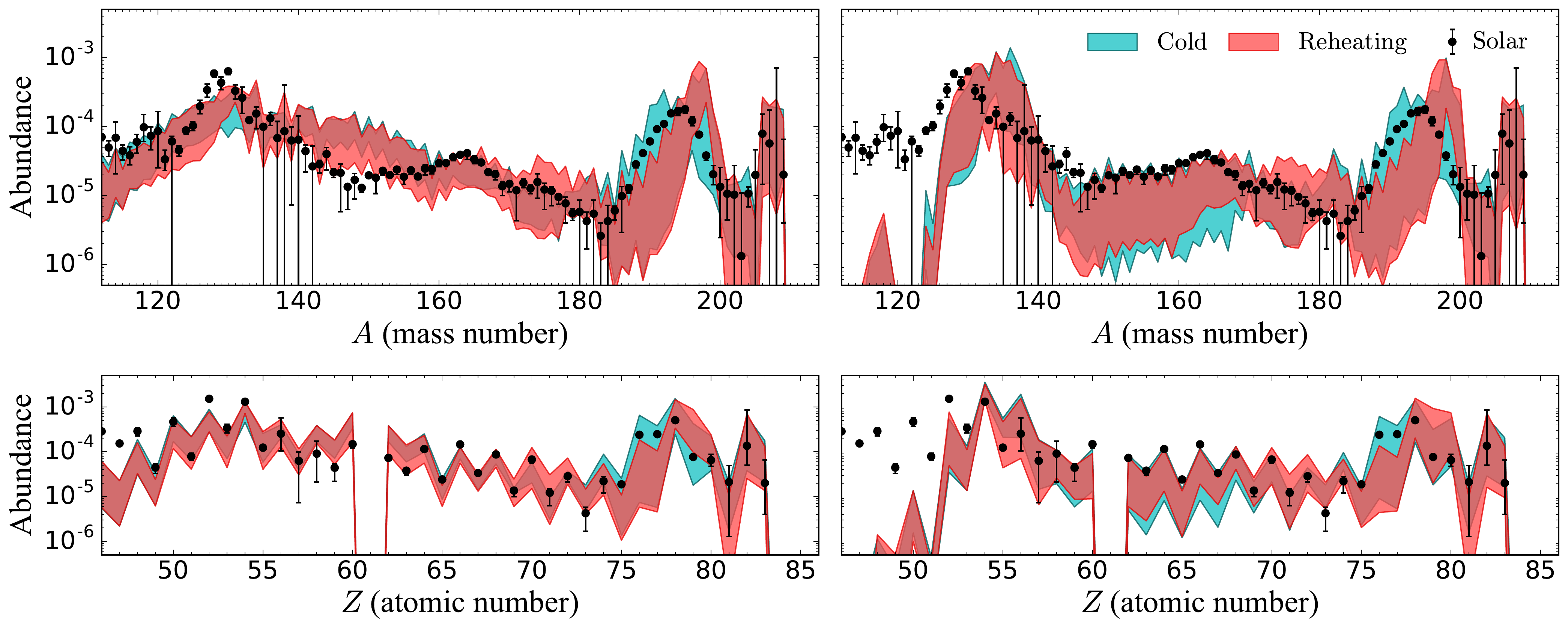}
\caption{The range in the calculated low-$Y_e$ r-process abundances as a function of mass number $A$ (upper row) and atomic number $Z$ (lower row) generated from the same ten mass models as in Fig.~\ref{fig:Yep27}, assuming \cite{kodama1975} (left panels) and a symmetric split (right panels) for the fission fragment distribution. Turquoise bands represent very neutron rich, cold merger outflow conditions (\citealt{Just2015}) without reheating while red bands represent conditions for ``slow" ejecta (\citealt{MT2015}) with reheating included. The dots are the observed solar r-process residuals \citep[taken from][]{arnould07}. \label{fig:abund}}
\end{figure*}

Since nuclear binding (mass) is critical input for other nuclear properties relevant to the r-process, we self-consistently update nuclear reactions \citep{mumpower2017} and decay modes \citep{mumpower2016} as in \cite{mumpower2015} using the Los Alamos suite of statistical Hauser-Feshbach codes \citep{kawano2016}. We also include $\beta$-delayed and neutron-induced fission rates calculated within this same self-consistent framework. To accommodate the flexibility in nuclear input data required for these self-consistent calculations, we use the nucleosynthesis code PRISM (Portable Routines for Integrated nucleoSynthesis Modeling, \citealt{mumpower2017}). This code easily allows for all nuclear data to be user defined and permits a straightforward hierarchy so that theoretical values are overwritten with experimental ones (e.g. from NUBASE \citealt{nubase16}), when available.

We first consider the astrophysical conditions of a wind scenario with $Y_e = 0.27$ previously found to produce the best fit to a ``blue" kilonova light curve \citep{wollaeger17,evans17}. The bands in Figure~\ref{fig:Yep27} demonstrate the range of abundance predictions when different mass models are considered. Even with an order of magnitude variation in the dynamical timescale (different band colors), this environment is not capable of generating a full r-process. Only the second peak r-process elements are reached and Eu is not produced.

Since the electromagnetic counterpart of GW170817 was found to be consistent with lanthanide production, we next turn to neutron-rich dynamical ejecta and consider two different astrophysical conditions. The first one is a ``cold" merger outflow condition, such as can be found in the tidal tail ejecta, with an entropy of $10\, k_B/$baryon and $Y_e = 0.05$ (\citealt{Just2015}). The second one is an astrophysical trajectory for ``slow" ejecta (\citealt{MT2015}) with a neutron-to-seed ratio of $R_{n/s}\sim10^3$, which takes into account the reheating due to nuclear reactions with a temperature rise from $\sim0.2$\,GK to $\sim1.5$\,GK in $\sim 60$\,ms for the case considered here.

\begin{deluxetable*}{ccccc}
\tablewidth{0pc}
\tablecaption{The mass fraction range for $^{151}$Eu, $^{153}$Eu, as well as the relative abundance range $(Y_\mathrm{max}-Y_\mathrm{min})/\overline{Y}$ for all stable europium isotopes. $Y_\mathrm{max}$, $Y_\mathrm{min}$, and $\overline{Y}$ are the maximum, minimum, and mean europium r-process abundance, respectively, calculated with the set of ten mass models outlined in Figure~\ref{fig:abund} (see Section~\ref{sec:uncertainties}). \label{tab:delEu}}
\tablehead{
\colhead{\multirow{2}{*}{Astrophysical Trajectory}} & \colhead{\multirow{2}{*}{Fission Fragment Distribution}} & \colhead{$^{151}$Eu Mass Fraction} & \colhead{$^{153}$Eu Mass Fraction} & \colhead{Relative} \\
 & & \colhead{[$10^{-3}$]} & \colhead{[$10^{-3}$]} & \colhead{Abundance Range} }
\startdata
Cold outflow (no reheating) & \cite{kodama1975} & $(5.01 - 11.7)$   & $(3.92 - 8.75)$ & $0.776$ \\
(\citealt{Just2015}) & Symmetric Split & $(0.083 - 2.65)$  &  $(0.12 - 2.84)$   &  3.239\vspace{0.03 in} \\
\hline \vspace{-0.075 in} \\
``Slow" ejecta with reheating &  \cite{kodama1975} & $(2.67 - 13.3)$  & $(1.89 - 9.62)$ & 1.568 \\
(\citealt{MT2015}) & Symmetric Split & $(0.19 - 2.09)$  & $(0.24 - 2.23)$  & 2.755 \\
\enddata
\end{deluxetable*}

The ``cold" trajectory dynamics are such that the temperature and density drop makes most of the r-process proceed under an interplay between neutron capture and $\beta$-decay, since photodissociation falls out of equilibrium early. However, a scenario with reheating extends the time in which $(n,\gamma)\leftrightarrow (\gamma,n)$ equilibrium persists allowing for more late-time neutron capture that shifts and narrows the third peak relative to the the results in ``cold" conditions (see the different colored bands in Figure~\ref{fig:abund}). The bands in Figure~\ref{fig:abund} demonstrate that the sensitivity of reaction rates to nuclear mass can alone produce an order of magnitude variation in the abundance for the considered dynamical ejecta trajectories.

Additional complications arise in these very neutron-rich environments since the r-process path proceeds to the region of the nuclear chart where fission reactions dominate, and different fission prescriptions lead to distinct abundance patterns. The left and right panels of Figure~\ref{fig:abund} compare abundance predictions using the fission fragment distribution of \cite{kodama1975} versus assuming a simple, symmetric split for the fissioning nucleus. The fission prescription sets the distribution of the second r-process peak and the left edge of the rare-earth peak. The predictions from all mass models follow the common trend determined by the fission fragment distribution. On average, variations due to the fission prescription can lead to an overproduction (left panel) or underproduction (right panel) of Eu ($Z=63$).

For GW170817, the constraints on the 2nd and 3rd peak r-process yields are set by the amount of high-opacity lanthanide elements needed to explain the late-time ``red" kilonova emission. Uncertainties in the nuclear cross-sections can produce the same total lanthanide ejecta, but vary the production of individual components wildly.  Keeping the amount of total lanthanides equal (their mass is constrained by the observations), we can study the additional uncertainty in the trans-lead, Eu, and r-process peak elements. The final column of Table~\ref{tab:yields} shows the yield range including the nuclear physics uncertainties outlined in Table~\ref{tab:delEu} which are based on the range of abundance predictions given by the nuclear mass models and fission fragment distributions outlined in this section. In this range of models, we find that the total Eu abundance can increase or decrease by a factor of 2 with the same total lanthanide abundance.

%%%%%%%%%%%%%%%%%%%%%%%%%%%
%%%%%%%%%%%%%%%%%%%%%%%%%%%
\section{Merger Rate Densities} \label{sect_methods}
%%%%%%%%%%%%%%%%%%%%%%%%%%%
%%%%%%%%%%%%%%%%%%%%%%%%%%%

Here we briefly describe the methodology outlined in C17a and used to connect population synthesis and GCE results to LIGO/Virgo's measurement.  We keep track of various sources of uncertainties in order to provide the confidence intervals of our results.
\\
\\
%%%%%%%%%%%%%%%%%%%%%%%%%%%%%
\subsection{Population Synthesis} \label{sect_pop_synth}
%%%%%%%%%%%%%%%%%%%%%%%%%%%%%

Population synthesis models predict NS-NS merger rates for stellar populations (e.g., \citealt{1999ApJ...526..152F,VossTauris03,MennekensVanbeveren14,2012ApJ...759...52D,2016Natur.534..512B,Chruslinska17}). Those models can be confronted with the observed merger rate estimated from several known NS-NS systems in the Milky Way ($21^{+28}_{-14}$ Myr$^{-1}$, \citealt{Kim15}). For comparison with other observational constraints such as short-duration gamma-ray bursts \citep{2014ARA&A..52...43B} and gravitational wave measurements \citep{2016ApJ...832L..21A}, 
a calculation of cosmological NS-NS merger rate densities is required. This involves tracing the formation of NS-NS progenitor systems according to the cosmic star formation history (CSFH, \citealt{2014ARA&A..52..415M}) and following their evolution until they merge using metallicity-dependent delay-time distributions (DTDs, see \citealt{2016ApJ...819..108B}).

The merger rate densities based on previous calculations \citep{2016Natur.534..512B} are too low compared to the latest LIGO/Virgo's estimates at low redshift\footnote{As a point of reference, NGC 4993, the host galaxy of GW170817, is at 40 Mpc ($z\sim0.01$).} ($1540^{+3200}_{-1220}$\,\gpy, \citealt{abbott17}). Those models have been revisited by \cite{Chruslinska17}.  For many realizations of the input physics (e.g., varied assumptions about the natal kicks, angular momentum loss, mass transfer), the classical evolution of isolated binaries typically leads to low merger rate densities at low redshifts ($\lesssim 50$\,\gpy). However, several models with specific common envelope physics, low angular momentum loss during Roche-lobe overflow, electron-capture SNe allowed in a wide range of initial stellar masses (with no natal kick applied), and reduced natal kicks for NS progenitors with heavily stripped envelopes, can produce local NS-NS merger rate densities as high as $\sim$\,$500-600$\,\gpy.

Uncertainties associated with the CSFH, the stellar initial mass function, the binary fraction, and the evolution of metallicity through cosmic time can further shift the predicted merger rate densities by a factor of $\sim$\,2. The highest merger rate density predicted with the calculations of \cite{Chruslinska17} is shown as the upper limit of the green shaded area in Figure~\ref{fig_GCE_LIGO_pop}. This limit ($\sim$\,10$^3$\,\gpy\ at redshift $z=0$) represents the most optimistic model increased by a factor of 2 to show the currently attainable maximum NS-NS merger rate density with population synthesis methods. The lower limit is the same as in C17a (but see \citealt{Chruslinska17}).

Since the DTD of NS-NS mergers typically follows a power-law in the form of $t^{-1}$ (see also \citealt{2012ApJ...759...52D}), close NS-NS binaries merge within a few Myrs after their formation. The evolution of the merger rate density should therefore follow the CSFH, which peaks at $z\sim2$. However, because NS-NS systems are most efficiently formed at high metallicities at $z<2$, the peak of the merger rate density is shifted to $z\sim1.5$.

%%%%%%%%%%%%%%%%%%%%%%%%%%%%%
\subsection{Galactic Chemical Evolution} \label{sect_gce}
%%%%%%%%%%%%%%%%%%%%%%%%%%%%%

We use a similar approach to calculate the merger rate densities required by GCE. However, instead of using the NS-NS merger rates predicted by population synthesis for individual stellar populations, we use the ones adopted in GCE simulations.  Those rates are calibrated to reproduce the [Eu/Fe]\footnote{[A/B]\,$=$\,log$_{10}$($n_A/n_B$)\,$-$\,log$_{10}$($n_A/n_B$)$_\odot$ where $n_A$ and $n_B$ are the number densities of elements $A$ and $B$. This elemental ratio is normalized to the solar value.} abundances observed in the Milky Way, assuming NS-NS mergers are the only source of r-process elements. We chose this particular elemental ratio for probing the r-process production because it represents the common observational target used in the seven GCE studies compiled and normalized in C17a.

The merger rate density required to reproduce the current [Eu/Fe] abundances in our Galaxy depends on the DTD of NS-NS mergers, on the chemical evolution code, and on the amount of Eu and Fe ejected by NS-NS mergers and supernovae, respectively (see Sections 5.3 and 7 in C17a for more details). The range of solutions for a DTD in the form of $t^{-1}$ is shown as the dark and light blue shaded areas in Figure~\ref{fig_GCE_LIGO_pop} (see Section~\ref{sect_r_MW} for details).

These rates, however, do not account for other sources of uncertainties such as the rate of Fe injection by Type~Ia supernovae (e.g. \citealt{2009A&A...501..531M,2016ApJ...824...82C,2017A&A...605A..59R}), the fraction of NS-NS binaries merging outside the star-forming region (e.g., \citealt{1999ApJ...526..152F,2014ApJ...792..123B,2017MNRAS.471.4488S}, but see \citealt{2016ApJ...829L..13B}), and the CSFH, which could all increase the width of the blue uncertainty bands.

%%%%%%%%%%%%%%%%%%%%%%%%%%%%%%
%%%%%%%%%%%%%%%%%%%%%%%%%%%%%%
\section{The r Process in the Milky Way} \label{sect_r_MW}
\begin{figure}
\center
\includegraphics[width=3.35in]{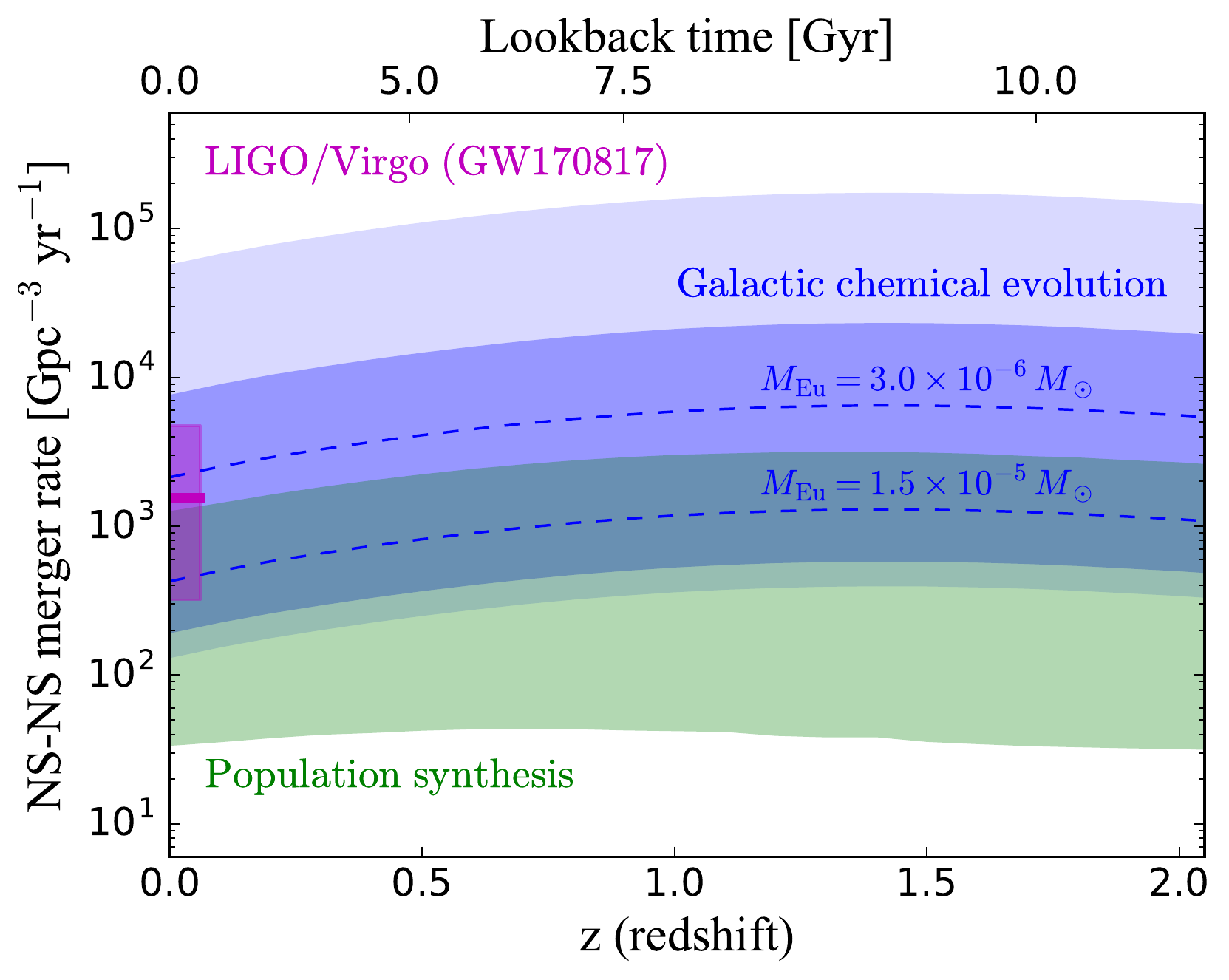}
\caption{Neutron star - neutron star (NS-NS) merger rate density as a function of redshift. The two blue dashed lines show the specific rates needed in galactic chemical evolution (GCE) studies to reproduce the amount of Eu observed in the Milky Way, when each NS-NS merger is assumed to eject on average $3\times10^{-6}$\,$M_\odot$ and $1.5\times10^{-5}$\,$M_\odot$ of Eu. These values represent the lower and upper limits of the total ejecta mass derived for GW170817 (Section~\ref{sect_yields}), when assuming a typical r-process abundance pattern for the ejecta. The dark blue shaded area shows the range of rates associated with those two values when GCE uncertainties are considered (see Section~\ref{sect_r_MW} for more details). The lighter (and larger) blue shaded area shows the range required when Eu yields are calculated theoretically from first principles, accounting for nuclear physics uncertainties (Section~\ref{sec:uncertainties}). The green shaded area represents the rates predicted using the population synthesis models of \cite{2016Natur.534..512B} and \cite{Chruslinska17}. The pink thick horizontal line and shaded area show the local rate and uncertainty provided by LIGO/Virgo from GW170817 (\citealt{abbott17}). Lookback times have been calculated using the cosmological parameters of \cite{2016A&A...594A..13P}. \label{fig_GCE_LIGO_pop}}
\end{figure}

%%%%%%%%%%%%%%%%%%%%%%%%%%%%%%
%%%%%%%%%%%%%%%%%%%%%%%%%%%%%%
Here we describe the implication of the first NS-NS merger detection on the chemical evolution of r-process elements in the Milky Way, assuming the frequency of NS-NS mergers per units of stellar mass formed is similar in different galaxies.
There is a degeneracy between the rate required by GCE and the average mass of Eu ejected by NS-NS mergers.  If NS-NS mergers release less r-process material, more mergers will be needed to recover the same level of enrichment (see also \citealt{2000ApJ...534L..67Q,2015NatPh..11.1042H,2018arXiv180101141H,2017arXiv171005805W}).  The two blue dashed lines in Figure~\ref{fig_GCE_LIGO_pop} show the merger rate densities required by GCE when the average Eu yields are $3\times10^{-6}$ and $1.5\times10^{-5}$\,$M_\odot$, representing the lower and upper limits derived for GW170817 (Section~\ref{sect_yields}) when assuming a typical r-process abundance pattern for the ejecta (e.g., \citealt{sneden08,arnould07,ji16}).

The dark blue shaded area surrounding these two lines represents the uncertainties caused by using different Fe yields for massive stars and by using different GCE studies to infer the required merger rate (see C17a). If we use theoretical calculations from first principles to calculate the abundance pattern of the ejecta, the range of Eu yields for GW170817 becomes significantly larger because of nuclear astrophysics uncertainties (see Section~\ref{sec:uncertainties}), which reduces our ability to constrain the contribution of NS-NS mergers using GCE arguments (lighter blue shaded area in Figure~\ref{fig_GCE_LIGO_pop}).

Overall, there is an overlap between GCE, population synthesis, and LIGO/Virgo between $\sim$\,300 and $\sim$\,1000\,\gpy. GCE and population synthesis are consistent with each other if NS-NS mergers eject on average $\gtrsim$\,$10^{-5}$\,$M_\odot$ of Eu.  The NS-NS merger rate densities derived from GW170817 (pink shaded area in Figure~\ref{fig_GCE_LIGO_pop}) 
are remarkably consistent with the GCE requirement if a typical r-process pattern is assumed for its ejecta (dark blue shaded area).  If GW170817 is statically a representative event, this detection suggests that NS-NS mergers are likely to be the main r-process site in the Milky Way and possibly in other galaxies.

Using the one-zone GCE code \texttt{OMEGA} \citep{2017ApJ...835..128C}, we calculate a current Galactic merger rate of $\sim$\,50 and $\sim$\,230\,Myr$^{-1}$ for Eu yields of $1.5\times10^{-5}$ and $3\times10^{-6}$\,$M_\odot$, respectively, when adopting the same input NS-NS merger prescription used to predicts merger rate densities in \gpy. The final ($z=0$) star formation rate in our Milky Way model is 2.5\,$M_\odot$\,yr$^{-1}$. Accounting for uncertainties in this final rate, which observationally ranges from 0.65 to 3\,$M_\odot$\,yr$^{-1}$ (e.g., \citealt{2010ApJ...710L..11R,2011AJ....142..197C,2015A&A...580A.126K}), and in the Fe yields used for massive stars in GCE studies (see Figure~1 in C17a), we obtain NS-NS merger rates in the range of [$5-100$] and [$35-495$]\,Myr$^{-1}$ for the upper and lower Eu yields limits, respectively.  

Those ranges are within the [$1-1000$]\,Myr$^{-1}$ range estimated by \cite{abadie10} from pulsar luminosities but are wider than the [$7-49$]\,Myr$^{-1}$ range derived by \cite{Kim15}. However, \cite{Chruslinska17} suggested that the range provided by \cite{Kim15} could be extended to [$2-210$]\,Myr$^{-1}$ if uncertainties in the pulsar luminosity function were included (see their Section~5.1), which significantly enlarges the overlap with our required Galactic rates of [$5-495$]\,Myr$^{-1}$.

To summarize, the GCE requirement overlaps and can be consistent with both the cosmic merger rate density in \gpy\ and the Galactic merger rate in Myr$^{-1}$.  In particular, our Galactic merger rates are in better agreement with \cite{Kim15} when the assumed Eu yields are $\gtrsim10^{-5}$\,$M_\odot$, which turns out to be the regime where GCE, population synthesis, and LIGO/Virgo are overlapping.

%%%%%%%%%%%%%%%%%%%%%%%%%%%%%%%%%%%%%%
%%%%%%%%%%%%%%%%%%%%%%%%%%%%%%%%%%%%%%
\section{Discussion} \label{sect_disc}
%%%%%%%%%%%%%%%%%%%%%%%%%%%%%%%%%%%%%%
%%%%%%%%%%%%%%%%%%%%%%%%%%%%%%%%%%%%%%
Here we compare our results and findings with other work, highlight a potential tension between population synthesis and GCE, and discuss the possibility that GW170817 is not a representative even in terms of its rate, total ejected mass, and abundance pattern.

%%%%%%%%%%%%%%%%%%%%%%%%%%%%%%%%%
\subsection{Analytical Estimates} \label{sect_disc_analytical}
%%%%%%%%%%%%%%%%%%%%%%%%%%%%%%%%%
Instead of using GCE simulations and reproducing stellar abundances, analytical calculations can also be used to estimate the role of NS-NS mergers on the Galactic r-process enrichment (e.g., \citealt{2010MNRAS.406.2650M,bauswein13a,2013MNRAS.430.2585R,2015NatPh..11.1042H}). Those calculations are based on the total mass of r-process elements $M_\mathrm{r,tot}$ currently present the Milky Way, which can be obtained by multiplying the total mass of baryons in our Galaxy by the mass fraction of the r-process in the solar system (\citealt{2000ApJ...534L..67Q}). This quantity can be divided by the lifetime of the Galaxy ($\sim$10\,Gyr) to calculate the average mass injection rate of r-process events. Then, by adopting different ejected masses for NS-NS mergers, one can infer the merger rates needed to recover $M_\mathrm{r,tot}$.

By combining the properties of GW170817 with analytical calculations similar to the ones described above, several studies have demonstrated that NS-NS mergers can synthesize enough r-process material to be the dominant site in the Milky Way (\citealt{abbott17c,cowperthwaite17,chornock17,2017arXiv171005442G,kasen17,rosswog17b,tanaka17b,2017arXiv171005805W,2018arXiv180101141H}). GCE simulations and analytical calculations are thus converging toward a similar message.  However, uncertainties in the mass ejected by NS-NS mergers are still significantly affecting the predictive power of both approaches.

%%%%%%%%%%%%%%%%%%%%%%%%%%%%%%%%%%%%%%%%%%%%%%%%%%
\subsection{Using Other Dynamical Ejecta Estimates} \label{disc_other_m_dyn}
%%%%%%%%%%%%%%%%%%%%%%%%%%%%%%%%%%%%%%%%%%%%%%%%%%
In this work, we used the conservative mass range of $0.002-0.01$\,$M_\odot$ for the dynamical ejecta of GW170817, which represents the low-$Y_e$ component that synthesizes Eu (see Section~\ref{sec:ejecta}). However, as seen in Table~\ref{tab:ejmasses}, some studies derived larger values that can reach up to $\sim$\,0.04\,$M_\odot$. Using this larger value yields 4 times more Eu compared to our upper limit of $1.5\times10^{-5}$\,$M_\odot$. In that case, the NS-NS merger rate density required by GCE drops to $\sim$100\,\gpy, which is about 3 times lower than the lower limit established by LIGO/Virgo.  Therefore, an upper limit of $0.01$\,$M_\odot$ for the dynamical ejecta seems to be more consistent with the current observed merger rate (see dark blue shaded area in Figure~\ref{fig_GCE_LIGO_pop}).

However, as discussed in Section~\ref{disc_unrep_event}, we cannot exclude the possibility that GW170817 is an unusual event, in which case the merger rate density and average mass ejected per NS-NS merger event would be subject to changes in the upcoming years.

%%%%%%%%%%%%%%%%%%%%%%%%%%%%%%%%%%%%%%%%%%%%%%%%%%
\subsection{Perspective of Population Synthesis Models} \label{disc_pop_synth}
%%%%%%%%%%%%%%%%%%%%%%%%%%%%%%%%%%%%%%%%%%%%%%%%%%
Although population synthesis models overlap with GCE and LIGO/Virgo (see green shaded area in Figure~\ref{fig_GCE_LIGO_pop}), their predicted merger rate densities tend to be lower. As a matter of fact, when addressing NS-NS merger rates in the context of old spheroidal galaxies with extinct star formation, similar to NGC~4993 which hosted GW170817, population synthesis models predict rates that are significantly lower than the ones established by LIGO/Virgo (\citealt{2017arXiv171200632B}). This discrepancy could be solved by improving the physics behind the formation of NS-NS mergers in theoretical calculations, or by favouring mergers with longer delay times. The latter option would, however, potentially be in tension with the chemical evolution trend of Eu in the Milky Way, which overall seems to behave like an alpha element with short production timescales (e.g., \citealt{2016A&A...586A..49B,2017arXiv171103643S}, see also discussion in C17a and \citealt{2018arXiv180101141H}).
\\
\\
%%%%%%%%%%%%%%%%%%%%%%%%%%%%%%%%%%%%%%%%%%%%%%%%%%
\subsection{In Case of a Non-Representative Event} \label{disc_unrep_event}
%%%%%%%%%%%%%%%%%%%%%%%%%%%%%%%%%%%%%%%%%%%%%%%%%%
So far, only one NS-NS has been detected by LIGO/Virgo, which means that GW170817 could be an unusual and non-representative event. The merger rate density could actually be lower, meaning that GW170817 has been detected earlier than statistically expected. If this is the case, the derived NS-NS merger rate will decrease as the LIGO/Virgo's observing time gets longer. From a GCE perspective, given the uncertainties, NS-NS mergers could still be the main r-process site even if the  merger rate density was reduced, as long as it does not drop below $\sim$\,$100-200$\,\gpy. On the other hand, if GW170817 has been detected later than statistically expected, the actual merger rate density could be higher.

GW170817 could also be unusual in terms of its total mass ejected. If NS-NS mergers eject on average more or less mass than GW170817, the range of merger rate densities required by GCE would be modified. Indeed, depending on the masses and mass ratio of the two neutron stars in the merger, the dynamical and wind ejecta masses can vary by a factor of $2-4$ \citep{korobkin12}.  The uncertain inferred neutron star mass ratios range from 0.4 to 1.0 \citep{abbott17}.  If the mass ratio is closer to 0.4, the ejected mass may be higher than representative values.

%%%%%%%%%%%%%%%%%%%%%%%%%%%%%%%%%%%%%%%%%%%%%%%%%%
\subsection{Using Europium as the r-Process Tracer} \label{disc_using_Eu}
%%%%%%%%%%%%%%%%%%%%%%%%%%%%%%%%%%%%%%%%%%%%%%%%%%
To connect GCE with LIGO/Virgo's detection, we used Eu as the r-process tracer to calibrate the required NS-NS merger rates.  We recall that in GCE simulations, only one r-process abundance pattern is typically applied to all NS-NS mergers. To define whether these events are frequent enough to explain the r-process in the Milky Way, the adopted abundance pattern needs to represent the average yields synthesize by NS-NS mergers. In that regard, it is justified to use the solar r-process residuals.

However, as mentioned in Section~\ref{sec:adopted_ab_pat}, it is not guarantee that the 3rd to 2nd r-process peak ratio ejected by GW170817 follows the solar distribution. In the case where the 2nd peak is overestimated relative to the 3rd peak, using Eu from Table~\ref{tab:yields} with GCE to quantify the role of NS-NS mergers would be irrelevant, as matching Eu would lead to an overestimation of the 2nd peak. On the other hand, if the 2nd peak in the GW170817 ejecta is underestimated, matching Eu with GCE would be reliable, but additional r-process sites would be needed to generate the \textit{missing} lighter r-process elements (e.g., neutrino-driven winds in CC~SNe, \citealt{2013JPhG...40a3201A}).

With only one confirmed merger event, it is difficult to established what is the typical r-process ejecta of NS-NS mergers. In case mergers can produce a variety of r-process abundance patterns, one would need to constrain the probability distributions of the ejecta before testing the contribution of NS-NS mergers using GCE simulations. Theoretical calculations could be use to define such distributions, but variations caused by nuclear physics uncertainties first need to be reduced. From an observational point of view, there are variations in the r-process abundance patterns of metal-poor stars when comparing with the light and heavy parts of the solar r-process composition (e.g., \citealt{sneden08,2010ApJ...724..975R,2014ApJ...797..123H}). Whether variations in NS-NS merger yields could participate in this observational feature needs further investigation.

As for GCE, we note that analytical calculations (see Section~\ref{sect_disc_analytical}) also currently rely on the solar r-process residuals to test whether NS-NS mergers are the dominant r-process site.

%%%%%%%%%%%%%%%%%%%%%%%%%%%%%%%%%%%%%%%%
\section{Conclusions} \label{sect_concl}
%%%%%%%%%%%%%%%%%%%%%%%%%%%%%%%%%%%%%%%%
We addressed the implication of the first NS-NS merger detected by LIGO/Virgo (GW170817) on the origin of r-process elements. Using the ejected yields estimated for GW170817 (see Table~\ref{tab:yields}), the range of merger rate densities of $320-4740$\,Gpc$^{-3}$\,yr$^{-1}$ derived by LIGO/Virgo is consistent with the range required by galactic chemical evolution (GCE) to explain the europium (Eu) abundances observed in the Milky Way, assuming NS-NS mergers are the dominant r-process site. Our results are based on a compilation recent GCE studies 
that used a wide variety of numerical approaches ranging from one-zone homogeneous models to cosmological hydrodynamic simulations.

If GW170817 is a representative event and has a typical r-process signature, this new gravitational wave detection supports the theory that NS-NS mergers are the dominant source of r-process elements (see Figure~\ref{fig_GCE_LIGO_pop}). In fact, if NS-NS mergers eject on average $\sim$\,$10^{-5}$\,$M_\odot$ of Eu, there is an overlap between GCE, population synthesis, Galactic merger rates, and LIGO/Virgo. In case GW170817 is an unusual event, the actual merger rate and typical ejecta mass could be different.  But even if the merger rate density is reduced to $\sim100-200$\,\gpy, NS-NS mergers could still be the dominant r-process site, as long as the typical Eu yields stay larger than $\sim$\,$10^{-5}$\,$M_\odot$.

We cannot exclude the possibility that the relative abundances produced by GW170817 differ from the solar r-process residuals. If that is the case, more NS-NS mergers need to be detected to better constrain the variety of abundance patterns associated with those events, and to be able to quantify their contribution using GCE simulations.  If such variety exists, determining whether NS-NS mergers are the dominant site of the r-process will require a multi-elemental analysis rather than a quantification solely based on Eu.

If nuclear network calculations are used instead of assuming a typical r-process pattern for GW170817, we found that uncertainties in nuclear masses and fission properties need to be reduced in order to better constrain the role of NS-NS mergers on the chemical evolution of r-process elements using LIGO/Virgo's detections. In any event, it is clear that significant advancements in our knowledge of the properties of nuclei far from stability are required to understand NS-NS merger nucleosynthesis from first principles.

%%%%%%%%%%
%%%%%%%%%%
\acknowledgments
%%%%%%%%%%
%%%%%%%%%%

We are thankful to the anonymous referee for providing constructive feedback that improved our manuscript.
This research is supported by the National Science Foundation (USA) under grant No. PHY-1430152 (JINA Center for the Evolution of the Elements) and by the ERC Consolidator Grant (Hungary) funding scheme (project RADIOSTAR, G.A. n. 724560). K.B. acknowledges support from the NCN grants Sonata Bis 2 (DEC-2012/07/E/ST9/01360), OPUS (2015/19/B/ST9/01099) and OPUS (2015/19/B/ST9/\\03188). M.M., N.V. and R.S. are supported by the U.S. Department of Energy under Contract No. DE-AC52-07NA27344 for the topical collaboration Fission In R-process Elements (FIRE). This work was also supported in part by the U.S. Department of Energy under grant number DE-SC0013039 (R.S. and T.S.). A portion of this work was also carried out under the auspices of the National Nuclear Security Administration of the U.S. Department of Energy at Los Alamos National Laboratory under Contract No. DE-AC52-06NA25396 (C.F., O.K., R.W., M.M.). This research used resources provided by the Los Alamos National Laboratory Institutional Computing Program (O.K.,R.W.).

\software{\texttt{OMEGA} \citep{2017ApJ...835..128C},
          \texttt{PRISM} \citep{mumpower2017},
          \texttt{StarTrack} \citep{2002ApJ...572..407B,2008ApJS..174..223B},
          {Los Alamos suite of statistical Hauser-Feshbach codes} \citep{kawano2016},
          \texttt{NumPy} \citep{2011arXiv1102.1523V},
          \texttt{matplotlib} (\url{https://matplotlib.org}).}

%\software{}


\vspace{5mm}
%\facilities{HST(STIS), Swift(XRT and UVOT), AAVSO, CTIO:1.3m,




\bibliographystyle{yahapj}
\bibliography{ms}



\end{document}